\documentclass[%
 reprint,
 amsmath,amssymb,
 aps,
]{revtex4-2}

\usepackage{graphicx}% Include figure files
\usepackage{dcolumn}% Align table columns on decimal point
\usepackage{bm}% bold math
\begin{document}

\preprint{APS/123-QED}

\title{Pushing the room temperature continous-wave operation limit of GaSb-based interband cascade lasers beyond 6\,\textmu m}%Increasing the long wavelength limit of GaSb-based interband cascade lasers in continuous-wave operation}% Force line breaks with \\
%\thanks{A footnote to the article title}%

\author{J. Nauschütz$^1$}
\author{H. Knötig$^2$}
\author{R. Weih$^1$}
\author{J. Scheuermann$^1$}
\author{J. Koeth$^1$}
\author{S. Höfling$^3$}
\author{B. Schwarz$^2$}%
\email{E-Mail: benedikt.schwarz@tuwien.ac.at}
\affiliation{ $^1$nanoplus Nanosystems and Technologies GmbH, Oberer Kirschberg 4, 97218 Gerbrunn, Germany\\
  $^2$Institute of Solid State Electronics, TU Wien, Gusshausstrasse 25-25a, 1040 Vienna, Austria\\
  $^3$Technische Physik, Physikalisches Institut, Universität Würzburg, Am Hubland, 97074 Würzburg, Germany}%

\date{\today}% It is always \today, today,
             %  but any date may be explicitly specified

\begin{abstract}
We present GaSb-based interband cascade lasers emitting at a center wavelength of 6.12\,\textmu m at 20°C in continuous-wave operation up to a maximum operating temperature of 40°C. Pulsed measurements based on broad area devices show improved performance by applying the recently published approach of adjusting the Ga$_{1-x}$In$_x$Sb layer thickness in the active region to reduce the valence intersubband absorption. The W-quantum well design adjustment and the optimization of the electron injector, to rebalance the electron and hole concentrations in the active quantum wells, improved the device performance, yielding room temperature current densities as low as 0.5\,kA/cm$^2$ for broad area devices under pulsed operation. As a direct result of this improvement the long wavelength limit for GaSb-based ICLs in continuous wave operation could be extended. For an epi-side down mounted 23\,\textmu m wide and 2\,mm long device with 9 active stages and high-reflectivity back facet the threshold power is below 1\,W and the optical output power is over 25\,mW at 20°C in continuous-wave mode. Such low-threshold and low-power consumption interband cascade lasers are especially attractive for mobile and compact sensing systems.
\end{abstract}

%Use showkeys class option if keyword
                              %display desired
\maketitle

%\tableofcontents

\section{Introduction}
In 1995 the basic concept of interband cascade lasers (ICLs) was proposed by Rui Q. Yang \cite{yang1995}, with the first experimental demonstration following in 1997 \cite{lin1997}. Since then ICLs have emerged as promising light sources for sensor applications in the spectral range between 3-6\,\textmu m \cite{meyer2020, vurgaftman2015} with a performance sweet spot between 3-4\,\textmu m. Devices in this range show superior performance, i.e., low threshold current densities, high operation temperatures and comparably high output power. The objective of this paper is to extend the long wavelength limit of GaSb-based ICLs in continuous-wave (cw) operation and improve their performance. An important milestone in ICL performance development was achieved in 2011, addressing the balancing of the internally generated charge carriers via a significant increase of the doping concentration in the electron injector \cite{vurgaftman2011}. Recently, a reduction of the valence intersubband absorption has been demonstrated by adjusting the Ga$_{1-x}$In$_x$Sb hole quantum well (h-QW) thickness, leading to further significant improvement in ICL performance \cite{knoetig2022}. For GaSb-based ICLs and room temperature (RT) operation, the longest published wavelengths are 6.8\,\textmu m for pulsed broad area devices \cite{bader2021} and 5.6\,\textmu m for narrow Fabry-Pérot (FP) lasers in cw operation \cite{bewley2012}. In the range of 2.8-5.2\,\textmu m \cite{scheuermann2017, edlinger2014} single mode operation was demonstrated at room temperature for devices employing distributed feedback (DFB) gratings. ICLs with wavelengths beyond 6.1\,\textmu m have so far been realized by plasmon-enhanced cladding layers on InAs substrates \cite{yang2019, rassel2017, dallner2013}. Broad area InAs-based ICLs emit up to 13.2\,\textmu m in pulsed operation at 120\,K \cite{massengale2022} and up to 11\,\textmu m at 97\,K in cw-operation \cite{li2015}. The wavelength region around 6\,\textmu m is contested by both ICLs and quantum cascade lasers (QCLs), that are also realized in this wavelength window~\cite{cheng2022, mawst2022, razeghi2003}. However, QCLs usually show higher threshold current densities and threshold voltages than ICLs. By extending the wavelength range of ICLs beyond 6\,\textmu m combined with the low power consumption, we envision that these light sources will successfully compete with QCLs and InAs-based ICLs in low-power applications. The extension of the emission wavelength paves the way towards compact and mobile analyzers that target further analytes such as NO$_2$ at 6.25\,\textmu m. 

\begin{figure*}
  \centering
  \includegraphics[width=\textwidth]{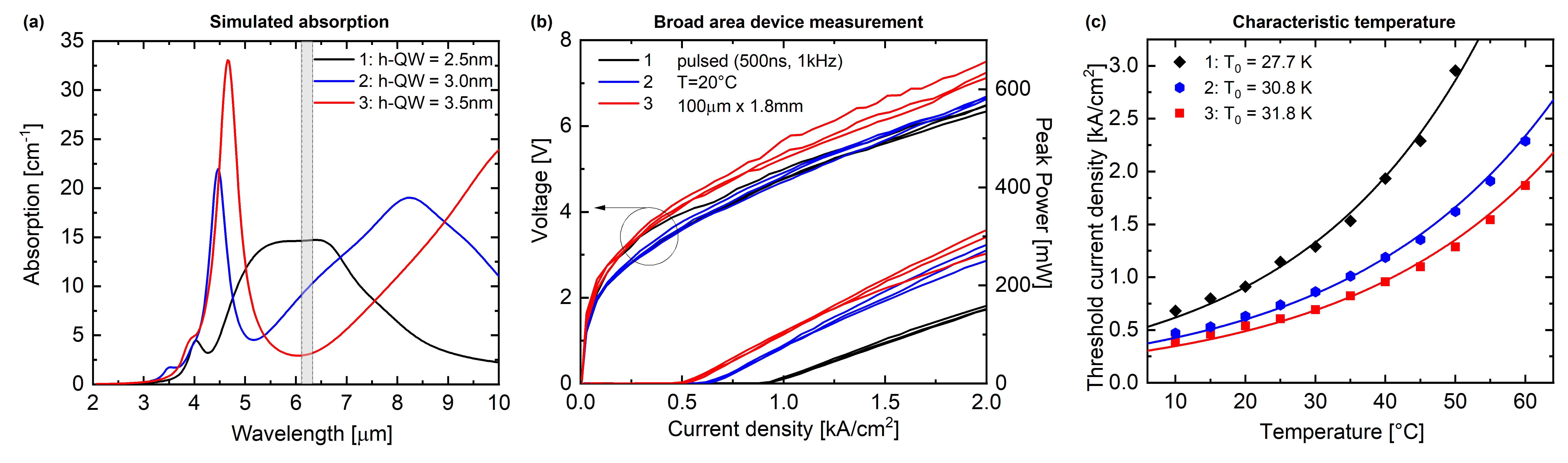}
  \caption{(a) Simulated absorption for sample~1: 2.5\,nm (black) / sample~2: 3.0\,nm (blue) / sample~3: 3.5\,nm (red) Ga$_{0.6}$In$_{0.4}$Sb  h-QW thicknesses and InAs layers of 2.77/2.31 (1), 2.71/2.26 (2), 2.68/2.24 (3). The examined wavelength range between 6.1 and 6.3\,\textmu m is marked. (b) Comparison of pulsed L-I-V curves for corresponding ICL structures 1-3 with h-QW thicknesses simulated in (a). The data is recorded at low duty cycle (1\,kHz, 500\,ns) and T=20°C with broad area devices fully etched through the active region. The devices are not soldered to the heat sink and the facets uncoated. The extracted performance parameters for the best device of each sample can be found in Table~\ref{tab:table1}. The higher threshold current and the low efficiency of sample 1 compared to sample 3 confirm the relationship reflected in the simulation in (a). (c) Temperature dependence of the threshold current density for the best broad area device plotted in (b) between 10°C and 60°C with exponential fit to J$_{th}$(T) and resulting characteristic temperatures T$_0$.  }
  \label{fig:fig1}
\end{figure*}

\section{Results and Discussion}
\label{sec:results}
In this work, we describe the three sequential steps we followed to cross the long wavelength limit for GaSb-based ICLs in cw-operation. As a first step and in accordance with the reported causative correlation between valence intersubband absorption and Ga$_{1-x}$In$_x$Sb h-QW thickness, we investigate the influence of h-QW thickness to reduce the absorption at longer wavelengths. In the second iteration, the electron injector of the active region was optimized by adjusting the doping concentration and shortening it by one InAs/AlSb pair. This stepwise approach is necessary because the valence intersubband absorption is expected to depend on the hole density. Since by varying the doping in the electron injector the hole density is changed in the active W-QW, different optimum values of the doping concentration are expected depending on the strength of the valence intersubband absorption. Therefore, the reduction of absorption needs to be done before the doping concentration can be optimized. Subsequently, the waveguide structure was improved by using thicker asymmetric separate confinement layers (SCLs) and thinner cladding layers. Finally, we were able to demonstrate cw operation at RT of an epi-side down mounted Fabry-Pérot laser with a central emission wavelength of 6.12\,\textmu m.

\subsection*{Reduction of intersubband absorption}
Significant performance improvement at an emission wavelength of 4.35\,\textmu m due to the minimization of valence intersubband absorption was demonstrated by Knötig et al.~\cite{knoetig2022}. This was achieved by a reduction of the thickness of the Ga$_{1-x}$In$_x$Sb layer in the active W-QW. Here, we investigate the performance of ICL structures at a wavelength of 6.15\,\textmu m. In contrast to the behavior at 4.35\,\textmu m, the simulations reveal a reduction of the absorption for wider h-QW thickness. Hence, our study serves two purposes: while tackling the improvement of ICL device performance beyond 6\,\textmu m, the relevance of the mitigation of valence intersubband absorption is validated for long wavelength devices. The expected reverse correlation of device performance and h-QW thickness at this wavelength offers the possibility to further explore the effect of valence intersubband absorption experimentally.
Following the experimental comparison at an emission wavelength of 4.35\,\textmu m, we investigate three h-QW layer thicknesses 2.5\,nm (1) / 3.0\,nm (2) / 3.5\,nm (3) in the first iteration presented  here. The simulated absorption in the W-QW for these three devices targeting an emission wavelength around 6.15\,\textmu m is shown in Fig.~\ref{fig:fig1}(a). Around the emission wavelength (depicted in gray), the widest h-QW (3 - red) should experience the lowest absorption and the narrowest (1 - black) the highest absorption. For a reasonable comparability of the structures, only the thicknesses of the two surrounding InAs layers were adjusted to set the emission wavelength (see section~\ref{sec:experimentaldetails}). All other parameters in the growth and fabrication of the devices are identical. Fig.~\ref{fig:fig1}(b) shows the pulsed light-current-voltage (L-I-V)  characteristics of the three best broad area devices based on ICL samples 1, 2 and 3 which were etched fully through the active region and were operated at 20°C and a duty cycle of  of 0.05\,\%.
\begin{table*}
 \caption{Comparison of laser characteristics of the investigated broad area ICL structures (best device performance): Ga$_{0.6}$In$_{0.4}$Sb h-QW thickness, Si-doping concentration in InAs wells of the electron injector, emission wavelength $\lambda$, threshold current density J$_{th}$, optical output power P$_{out}$ at 2\,kA/cm$^2$ per facet, efficiencies near threshold $\eta_{slope}$ and characteristic temperature T$_{0}$. Performance parameters were measured on 100\,\textmu m x 1.8\,mm devices, fully etched through the active region, at  20°C and low duty cycle of 0.05\,\%. %duty cycle DC for measured J$_{th}$, P$_{out}$, $\eta_{slope}$ and device dimensions.
 }
  \begin{ruledtabular}
  \renewcommand{\arraystretch}{1.25}
  \begin{tabular}{cccccccc}
      Sample   & h-QW    & Doping & $\lambda$  & J$_{th}$ & P$_{out}$ & $\eta_{slope}$ & T$_{0}$ \\ % & DC & Dimensions\\
               & [nm] &  [cm$^{-3}$] & [\textmu m] & [kA/cm$^2$] & [mW] & [mW/A] & [K] \\ \hline % & [\%]  & [\textmu m x mm] \\ 
               
    1 & 2.5  & 3E18 & 6.18  & 0.92 & 159 & 150  & 27.7 \\ % & 0.05 & 100 x 1.8    \\
    2 & 3.0  & 3E18 & 6.33  & 0.63 & 284 & 197  & 30.8 \\ % & 0.05 & 100 x 1.8   \\
    3 & 3.5  & 3E18 & 6.12  & 0.53 & 314 & 200  & 31.8 \\ \hline% & 0.05 & 100 x 1.8    \\
   
    4 & 3.0  & 1E18 & 6.25  & 0.50 & 319 & 220 & 35.9 \\ % & 0.05 & 100 x 1.8    \\
    5 & 3.5  & 1E18 & 6.21  & 0.52 & 390 & 262 & 40.2 \\ % & 0.05 & 100 x 1.8    \\
%    \midrule
%    6 & 3.5  & 1E18 & 6.12  & 0.61 & 30 (@1.4) & 55  &  40.2 & 100 & 23 x 2.0    \\
%    \bottomrule
  \end{tabular}
  \end{ruledtabular}
  \label{tab:table1}
\end{table*}
The simulated correlation between the performance-deteriorating absorption and the h-QW thickness is unequivocally confirmed by the measurements. Significant differences in the measured device performance, like threshold current density J$_{th}$, output power P$_{out}$, and efficiency $\eta_{slope}$, are observed, with the optimized design 3 showing the best performance. The performance characteristics of the best broad area device of each design are summarized in Table~\ref{tab:table1}.
\begin{figure*}
  \centering
  \includegraphics[width=\textwidth]{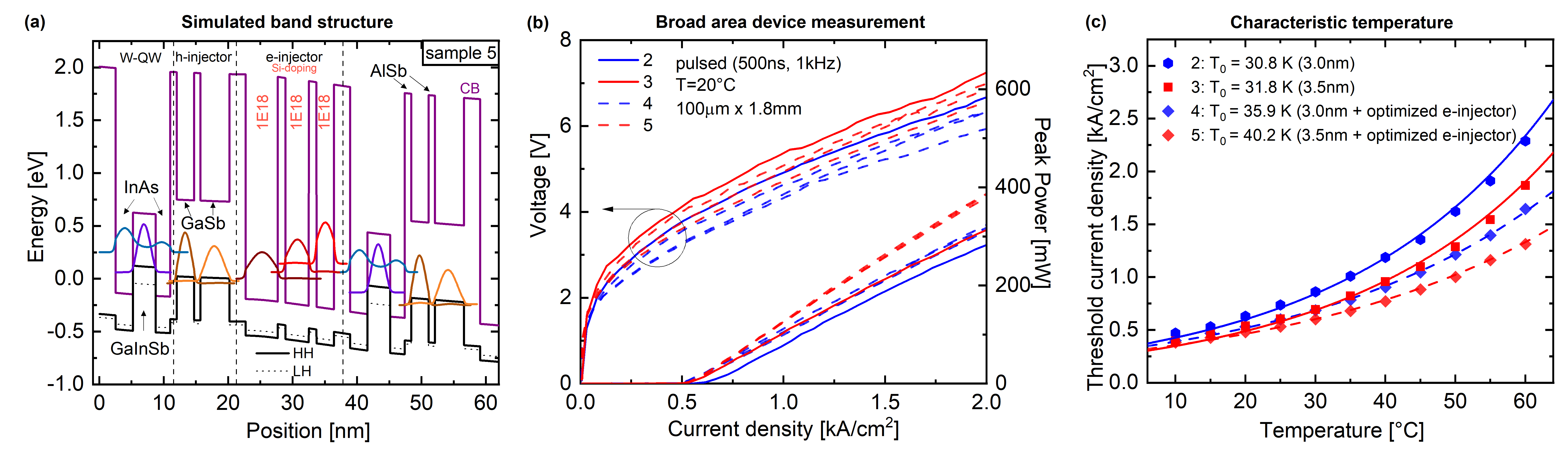}
  \caption{(a) Band structure of sample 5 with probability of presence for electrons and holes in 1.5 stages of the active region under an applied electric field of 55\,kV/cm for proper alignment of energy levels (conduction band (CB) - purple; valence band (VB) - black (heavy/light hole - solid/dashed). In contrast to samples 1-3, the electron injector is shortened and the doping is reduced to balance the internal carrier densities. (b) Comparison of pulsed L-I-V curves for corresponding ICL structures 2-5 at low duty cycle and T=20°C. Samples 2 (blue) and 4 (dashed blue) have an h-QW thickness of 3.0\,nm while samples 3 (red) and 5 have a 3.5\,nm (dashed red) thick h-QW. The electron injector optimization of samples 4 and 5 improves the performance of the broad area devices attributed to more adequate balancing of charge carrier densities in the W-QW. (c) Measured threshold current densities for sample 2-5 at different temperatures (10°C - 60°C) and corresponding exponential fits to extract the characteristic temperatures T$_0$ with (dashed) and without (solid) electron injector optimization at two different h-QW thicknesses.}
  \label{fig:fig2}
\end{figure*}
The best devices of sample 3 in comparison to sample 1 show a 42\,\% decrease in threshold current density and a doubling of optical output power from 159\,mW to 314\,mW at 2\,kA/cm$^2$. The difference between performance parameters of samples 2 and 3 is smaller, but also reflects the simulated relationship between h-QW thickness and absorption. It should also be mentioned that the optical output power varies from device to device at high current densities. In Ref.~\cite{canedy2017} threshold current densities of 0.66-0.80\,kA/cm$^2$ are reported for broad area lasers at a wavelength of 6.1\,\textmu m at 25°C. With an h-QW width of 3.5\,nm and thus reduced W-QW absorption, an even lower threshold current density of 0.60\,kA/cm$^2$ (sample 3 - red) was measured at 25°C. The temperature dependence of the threshold current densities for samples 1, 2 and 3 was fitted based on $J_{th}(T) = J_0 \cdot exp(T/T_0)$ and is plotted in Fig.~\ref{fig:fig1}(c). Measured between 10°C and 60°C, the results demonstrate that sample 3 exhibits the lowest threshold current densities of all three samples~1-3 over the entire temperature range. Moreover, a stronger increase in threshold current density with temperature is observed for sample 1 compared to samples 2 and 3. This behavior fits the simulation result, showing the highest valence intersubband absorption for sample 1. The absorption is expected to depend on the hole density and consequently on the temperature, which is reflected in T$_0$. Compared to the published T$_0$ values from \cite{knoetig2022}, the absolute T$_0$ values here are smaller. It should be noted that the broad area devices in this work are completely etched through the active region to reduce current spreading. Thus, a possible explanation for the difference between T$_0$ values from deeply-etched and shallowly-etched devices could be a temperature-dependent current spreading.

\subsection*{Optimization of electron injector}
The following adjustments of the electron injector aim to more accurately balance the disequilibrium of carrier densities in the W-QW and thereby also reduce loss mechanisms such as Auger recombination \cite{vurgaftman2011}. The optimization of the electron injector is implemented after the reduction of the intersubband absorption, due to an expected shift of the optimum value of the doping depending on the presence of this absorption, which depends on the hole density.\\
\begin{figure*}
  \centering
  \includegraphics[width=\textwidth]{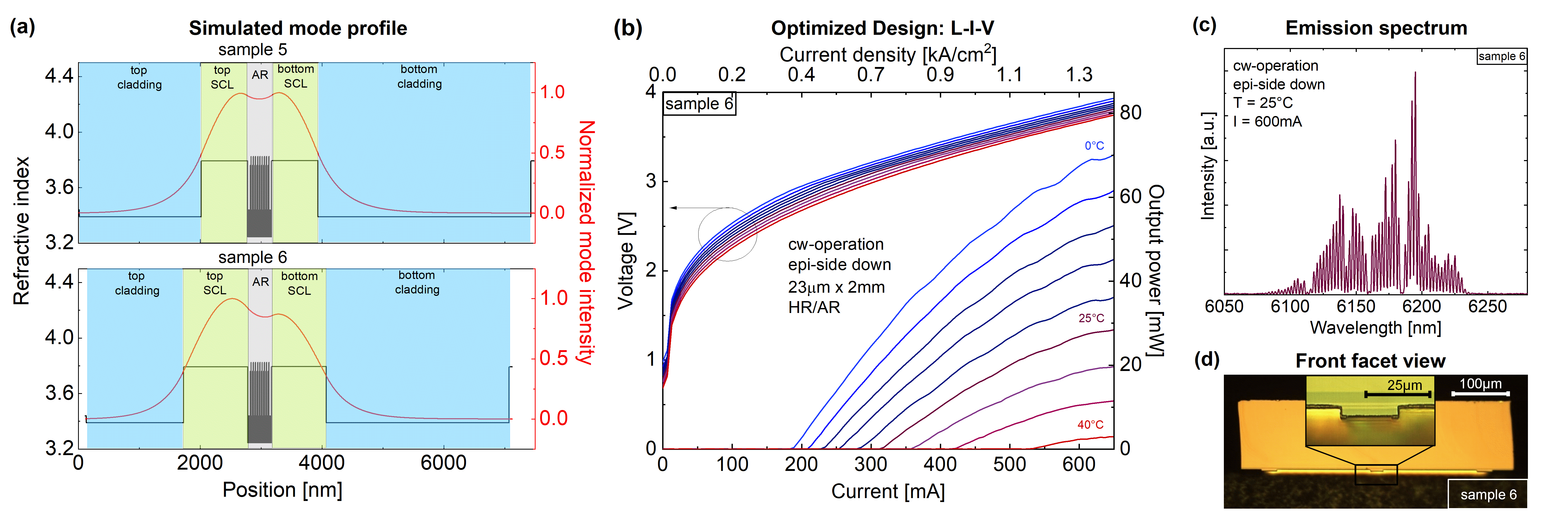}
  \caption{(a) Simulation of the vertical field distribution of the optical mode resulting from the refractive index profiles of samples~5 (top) and 6 (bottom). The waveguide design was optimized with asymmetrically thicker SCLs (green) and thinner cladding layers (blue). The normalized mode intensity (red) and the refractive index profile (black) are shown. (b) L-I-V curves of an epi-side down mounted FP ICL (23\,\textmu m wide and 2\,mm long - HR back facet / AR front facet) with epitaxy design~6 in cw-mode in steps of 5°C from 0°C to 40°C. At 0°C a threshold current density of 420\,A/cm$^2$ and optical output power of 70\,mW is achieved, which correspond to a wallplug efficiency of 3.2\,\%. The small dips in the power curve originate from absorption of water vapor in ambient air. (c) Emission spectrum of the epi-side down ICL with a central emission wavelength of 6.17\,\textmu m at 25°C and 600\,mA in cw-operation. (d) Microscope image of the front facet of this epi-side down ICL soldered on AlN heatspreader with AuSn. In the close-up of the ridge waveguide, the epitaxial structure of the waveguide (SCLs = light yellow; cladding layers = light blue) is visible.}
  \label{fig:fig3}
\end{figure*}
Since samples 2 and 3 with h-QW thicknesses of 3.0\,nm and 3.5\,nm show lower intersubband absorption than sample 1, the effects of electron injector changes were further examined for these two h-QW thicknesses. For samples 4 and 5 the length of the electron injector was reduced by one InAs/AlSb pair and the layer thicknesses were slightly adjusted. The band structure of the active region for sample 5 is shown in Fig. \ref{fig:fig2}(a). In addition, a main difference between the designs 2/3 and 4/5 is the doping of the electron injector. In samples 4 and 5 the Si-doping concentration in the electron injector was lowered lowered from 3$\cdot 10^{18}$ cm$^{-3}$ to 1$\cdot 10^{18}$ cm$^{-3}$. % In addition, the shorter electron injector could be advantageous to shorten the active zone and thus slightly increase the optical modal overlap and consequently the modal gain.
% is it easy to calculate the carrier densities?  
The L-I-V curves of the three best broad area devices of samples 4 and 5 (dashed) are plotted in Fig. \ref{fig:fig2}(b) in comparison to the best devices from samples 2 and 3 (solid). Due to the changes in the electron injector, an improvement in performance can be observed for both h-QW thicknesses. In the case of 3.0\,nm with optimized electron injector (sample 4), the performance can be improved to the point where it shows comparable threshold current densities, efficiencies, and output powers as sample 3, which has a 3.5\,nm wide h-QW and a longer electron injector with higher doping. For 3.5\,nm, no further reduction of the threshold current density is observed for the adapted electron injector design. However, the design change results in further increased efficiency and optical output power. Threshold current density was also investigated over a temperature range of 10°C to 60°C for sample 4 and 5 and is depicted in comparison to sample 2 and 3 in Fig.~\ref{fig:fig2}(c). Throughout the temperature range, a decrease in threshold current density due to the changes of the electron injector is observed for both 3.0\,nm h-QW thickness, sample 2 (blue) to sample 4 (dashed blue), and 3.5\,nm h-QW thickness, sample 3 (red) to sample 5 (dashed red). In the low temperature range up to 25°C, samples 3, 4 and 5 hardly differ. From higher temperatures of approx. 40°C onwards, sample 4 with optimized e-injector displays a slightly lower threshold current density than sample 3 without optimized injector.  In general, sample 5 with optimized h-QW thickness of 3.5\,nm and improved e-injector shows the lowest threshold current density over the entire temperature range and hence the highest characteristic temperature T$_0$ of 40.2\,K of samples 1-5.

\subsection*{Improved waveguide design}
As sample 5 shows the best potential for RT-cw-operation due to the highest characteristic temperature, we regrew the active region of sample 5 with an optimized waveguide design (sample 6), see Fig.~\ref{fig:fig3}(a). The widths of the cladding layers were reduced and the thickness of the SCLs was asymmetrically increased with a thicker top SCL. This approach is especially beneficial for thermal management, but also for minimization of internal material specific losses, as the GaSb SCLs show higher thermal conductivity and lower internal losses than the cladding layers \cite{vurgaftman2011ieee}. The thinner top cladding layer is particularly advantageous for subsequent epi-down mounting, as it allows superior heat dissipation from the device. Moreover the changes in the waveguide design shift the modal overlap from the cladding layers and active region to the SCLs. Comparing the changed waveguide design of sample 6 to sample 5, the optical overlap increases by 8\,\% in the SCLs, and decreases by 6\,\% in the cladding layers and by 2\,\% in the active region according to waveguide simulations. This adjustment of the layer thicknesses reflects a trade-off between the modal gain and the internal losses of the laser structure \cite{meyer2020}. Other limiting factors that determine the thickness of the cladding layers are the overlap of the optical mode with the contact metalization and the substrate, which should be small to minimize optical losses. And secondly, the appearance of leakage modes, which cause periodic losses and can occur if the lower cladding layer is chosen too thin \cite{bewley2004}.\\
In pulsed operation, the deeply etched broad area device shows the same characteristic temperature of 40.2\,K as sample 5 (see Fig.~\ref{fig:fig2}(c)). This shows a high reproducibility of the growth, since sample 5 and 6 feature the same active region and no further influence on the characteristic temperature due to the changes of the waveguide design were expected at low duty-cycle.\\ 
For performance tests in cw opertaion, FP lasers (23\,\textmu m wide and 2\,mm long) with high reflective (HR) / anti-reflective (AR) coating were fabricated from samples 5 and 6 and mounted epi-side down. The comparison of the cw-performance parameters shows an improvement in efficiency by more than~50\,\% from 96\,mW/A (sample 5) to 145\,mW/A (sample 6) at 20°C due to the changes in the waveguide structure. The temperature dependent L-I-V characteristics of the device from sample 6 are shown in Fig. \ref{fig:fig3}(b) with a low threshold power of only 0.71\,W at 20°C. Output powers of 70\,mW in cw operation at 0°C were observed, which corresponds to a wallplug efficiency of 3.2\,\%. The cw-emission spectrum at 25°C with a central emission wavelength of 6.17\,\textmu m is plotted in Fig. \ref{fig:fig3}(c) and a close-up of the front facet in Fig. \ref{fig:fig3}(d).

\section{Conclusion}
In this paper, the influence of the active Ga$_{1-x}$In$_x$Sb layer thickness on the device performance for ICLs emitting at 6.1\,\textmu m was studied experimentally. Performance improvements by adjusting the h-QW layer thickness reveal the relationship between valence intersubband absorption and Ga$_{1-x}$In$_x$Sb layer thickness in this wavelength range. This unequivocally proves that the performance improvement when changing the h-QW thickness is indeed due to the avoidance of intersubband resonances in the valence band, as at different wavelengths (4.35\,\textmu m and 6.1\,\textmu m) the opposite trend is observed. Hence, the experimental observations nicely match the theoretical results. Moreover, other changes in the band structure, i.e., different energy spacing between the hole injector level and the lower lasing level, when varying the h-QW thickness can be excluded as explanations, since these would lead to the same observed behavior for both wavelengths. Broad area devices with reduced absorption show 42\,\% decrease in threshold current density to a record value of 0.5 kA/cm$^2$ and a doubling of optical output power in comparison to devices with non adapted h-QW thicknesses. Also the design optimization in the electron injector is supported by performance improvements of broad area devices. After reducing the intersubband absorption by adjusting the h-QW thickness, the electron injector was optimized to further balance the carrier densities in the W-QW by using a significantly lower doping concentration and shortening the electron injector. In the last step, the waveguide design was optimized with regard to thermal management and loss minimization. And as a result of this study, laser emission over 6.1\,\textmu m with GaSb-based interband cascade lasers was demonstrated in cw-operation at RT for an epi-side down mounted device with HR back facet. With the presented design strategies, a threshold power below 1\,W is achieved with a ridge width of 23\,\textmu m, providing a solid basis for future GaSb-based DFB ICLs with excellent device performance at and beyond 6\,\textmu m. 
%Even though the width of the device with 23\,\textmu m is not yet optimized for low threshold power, a threshold power of less than 1\,W at room temperature is achieved. These optimizations of the epitaxial structure could therefore provide a solid basis for future GaSb-based ICLs with DFB grating with excellent device performance at 6.1\,\textmu m.

\section{Experimental details}
\label{sec:experimentaldetails}
All six ICL designs investigated were grown on a Te-doped GaSb substrate by molecular beam epitaxy with 9 active stages, see Table~\ref{tab:table1}. In the first three structures 1, 2, 3, the layer thicknesses of the W-QW were varied as follows: (1) 2.77\,nm InAs / 2.5\,nm Ga$_{0.6}$In$_{0.4}$Sb / 2.31\,nm InAs; (2) 2.71\,nm InAs / 3.0\,nm Ga$_{0.6}$In$_{0.4}$Sb / 2.26\,nm InAs and (3) 2.68\,nm InAs / 3.5\,nm Ga$_{0.6}$In$_{0.4}$Sb / 2.24\,nm InAs. The thicknesses of the InAs layers were adapted with the aim of achieving an emission wavelength of 6.14\,\textmu m. The electron injector region was grown with four InAs/AlSb layer pairs, with the three InAs wells closest to the W-QW doped with a Si-concentration of $3\cdot 10^{18}$\,cm$^{-3}$. \\
The other three structures 4, 5, 6 were grown with a W-QW thickness of 3.0\,nm (4) and 2.68\,nm / 2.24\,nm InAs layers and 3.5\,nm (5,6) and 2.72\,nm / 2.27\,nm InAs layers, respectively and also with a Ga$_{1-x}$In$_x$Sb composition of $x=0.4$. Moreover, the electron injector consists of only three InAs/AlSb pairs, where all three InAs wells were doped with a lower Si-concentration of $1\cdot 10^{18}$\,cm$^{-3}$.\\
All ICL structures except for sample 6 were grown with 760\,nm thick SCLs symmetrically arranged around the active stages, while the cladding layers have thicknesses of 2\,\textmu m (top) and 3.5\,\textmu m (bottom). Sample 6 has reduced cladding layer thicknesses of 1.6\,\textmu m (top) / 3.0\,\textmu m (bottom) and asymmetrically thicker SCLs with thicknesses of 1050\,nm (top) / 900\,nm (bottom).\\
For basic characterization of the different ICL structures~1-6, 100\,\textmu m wide and 1.8\,mm long broad area lasers were fabricated. The ridge waveguides were defined by optical lithography and wet etched through the active region using a phosphoric etching solution. The etch depths all lie between 3.8\,\textmu m and 4.0\,\textmu m, reaching the lower cladding layer. The metallic top contact serves as an etch mask and the backside is coated with an AuGe/Ni/Au contact. The broad area devices were characterized in pulsed mode (1\,kHz, 500\,ns). The devices are not mounted and the facets are left uncoated.\\
Ridge waveguides (RWGs) were fabricated from samples 5 and 6 using e-beam lithography and reactive ion etching. Si$_3$N$_4$ and SiO$_2$ passivation layers were deposited and for electrical contacting, the passivation on top of the ridges was removed by reactive ion etching. A Ti-Pt-Au contact was deposited and patterned via a lift-off process. In addition, a 6\,\textmu m thick Au layer was electroplated on top. The GaSb substrate was then thinned to 150\,\textmu m and coated with an AuGe/Ni/Au backside contact. After cleaving, all lasers were coated with an HR metal layer on the backside and Al$_2$O$_3$ AR layer on the front side. 23\,\textmu m wide and 2.0\,mm long FP lasers were mounted epi-side down with AuSn solder on an AlN heatspreader for efficient heat extraction from the device. Subsequently, the heatspreader was soldered onto a copper heat sink with InSn and characterized in cw-mode.

\section*{Acknowledgments}
J. N. and H. K. contributed equally to this work.


\begin{thebibliography}{21}%
	\makeatletter
	\providecommand \@ifxundefined [1]{%
		\@ifx{#1\undefined}
	}%
	\providecommand \@ifnum [1]{%
		\ifnum #1\expandafter \@firstoftwo
		\else \expandafter \@secondoftwo
		\fi
	}%
	\providecommand \@ifx [1]{%
		\ifx #1\expandafter \@firstoftwo
		\else \expandafter \@secondoftwo
		\fi
	}%
	\providecommand \natexlab [1]{#1}%
	\providecommand \enquote  [1]{``#1''}%
	\providecommand \bibnamefont  [1]{#1}%
	\providecommand \bibfnamefont [1]{#1}%
	\providecommand \citenamefont [1]{#1}%
	\providecommand \href@noop [0]{\@secondoftwo}%
	\providecommand \href [0]{\begingroup \@sanitize@url \@href}%
	\providecommand \@href[1]{\@@startlink{#1}\@@href}%
	\providecommand \@@href[1]{\endgroup#1\@@endlink}%
	\providecommand \@sanitize@url [0]{\catcode `\\12\catcode `\$12\catcode
		`\&12\catcode `\#12\catcode `\^12\catcode `\_12\catcode `\%12\relax}%
	\providecommand \@@startlink[1]{}%
	\providecommand \@@endlink[0]{}%
	\providecommand \url  [0]{\begingroup\@sanitize@url \@url }%
	\providecommand \@url [1]{\endgroup\@href {#1}{\urlprefix }}%
	\providecommand \urlprefix  [0]{URL }%
	\providecommand \Eprint [0]{\href }%
	\providecommand \doibase [0]{https://doi.org/}%
	\providecommand \selectlanguage [0]{\@gobble}%
	\providecommand \bibinfo  [0]{\@secondoftwo}%
	\providecommand \bibfield  [0]{\@secondoftwo}%
	\providecommand \translation [1]{[#1]}%
	\providecommand \BibitemOpen [0]{}%
	\providecommand \bibitemStop [0]{}%
	\providecommand \bibitemNoStop [0]{.\EOS\space}%
	\providecommand \EOS [0]{\spacefactor3000\relax}%
	\providecommand \BibitemShut  [1]{\csname bibitem#1\endcsname}%
	\let\auto@bib@innerbib\@empty
	%</preamble>
	\bibitem [{\citenamefont {Yang}(1995)}]{yang1995}%
	\BibitemOpen
	\bibfield  {author} {\bibinfo {author} {\bibfnamefont {R.~Q.}\ \bibnamefont
			{Yang}},\ }\bibfield  {title} {\bibinfo {title} {Infrared laser based on
			intersubband transitions in quantum wells},\ }\href
	{https://doi.org/https://doi.org/10.1006/spmi.1995.1017} {\bibfield
		{journal} {\bibinfo  {journal} {Superlattices and Microstructures}\ }\textbf
		{\bibinfo {volume} {17}},\ \bibinfo {pages} {77} (\bibinfo {year}
		{1995})}\BibitemShut {NoStop}%
	\bibitem [{\citenamefont {Lin}\ \emph {et~al.}(1997)\citenamefont {Lin},
		\citenamefont {Yang}, \citenamefont {Zhang}, \citenamefont {Murry},
		\citenamefont {Pei}, \citenamefont {Allerman},\ and\ \citenamefont
		{Kurtz}}]{lin1997}%
	\BibitemOpen
	\bibfield  {author} {\bibinfo {author} {\bibfnamefont {C.-H.}\ \bibnamefont
			{Lin}}, \bibinfo {author} {\bibfnamefont {R.~Q.}\ \bibnamefont {Yang}},
		\bibinfo {author} {\bibfnamefont {D.}~\bibnamefont {Zhang}}, \bibinfo
		{author} {\bibfnamefont {S.}~\bibnamefont {Murry}}, \bibinfo {author}
		{\bibfnamefont {S.}~\bibnamefont {Pei}}, \bibinfo {author} {\bibfnamefont
			{A.}~\bibnamefont {Allerman}},\ and\ \bibinfo {author} {\bibfnamefont
			{S.}~\bibnamefont {Kurtz}},\ }\bibfield  {title} {\bibinfo {title} {Type-ii
			interband quantum cascade laser at 3.8\,\textmu m},\ }\href@noop {}
	{\bibfield  {journal} {\bibinfo  {journal} {Electronics Letters}\ }\textbf
		{\bibinfo {volume} {33}},\ \bibinfo {pages} {598} (\bibinfo {year}
		{1997})}\BibitemShut {NoStop}%
	\bibitem [{\citenamefont {Meyer}\ \emph {et~al.}(2020)\citenamefont {Meyer},
		\citenamefont {Bewley}, \citenamefont {Canedy}, \citenamefont {Kim},
		\citenamefont {Kim}, \citenamefont {Merritt},\ and\ \citenamefont
		{Vurgaftman}}]{meyer2020}%
	\BibitemOpen
	\bibfield  {author} {\bibinfo {author} {\bibfnamefont {J.~R.}\ \bibnamefont
			{Meyer}}, \bibinfo {author} {\bibfnamefont {W.~W.}\ \bibnamefont {Bewley}},
		\bibinfo {author} {\bibfnamefont {C.~L.}\ \bibnamefont {Canedy}}, \bibinfo
		{author} {\bibfnamefont {C.~S.}\ \bibnamefont {Kim}}, \bibinfo {author}
		{\bibfnamefont {M.}~\bibnamefont {Kim}}, \bibinfo {author} {\bibfnamefont
			{C.~D.}\ \bibnamefont {Merritt}},\ and\ \bibinfo {author} {\bibfnamefont
			{I.}~\bibnamefont {Vurgaftman}},\ }\bibfield  {title} {\bibinfo {title} {The
			interband cascade laser},\ }\bibfield  {journal} {\bibinfo  {journal}
		{Photonics}\ }\textbf {\bibinfo {volume} {7}},\ \href
	{https://doi.org/10.3390/photonics7030075} {10.3390/photonics7030075}
	(\bibinfo {year} {2020})\BibitemShut {NoStop}%
	\bibitem [{\citenamefont {Vurgaftman}\ \emph {et~al.}(2015)\citenamefont
		{Vurgaftman}, \citenamefont {Weih}, \citenamefont {Kamp}, \citenamefont
		{Meyer}, \citenamefont {Canedy}, \citenamefont {Kim}, \citenamefont {Kim},
		\citenamefont {Bewley}, \citenamefont {Merritt}, \citenamefont {Abell},\ and\
		\citenamefont {Höfling}}]{vurgaftman2015}%
	\BibitemOpen
	\bibfield  {author} {\bibinfo {author} {\bibfnamefont {I.}~\bibnamefont
			{Vurgaftman}}, \bibinfo {author} {\bibfnamefont {R.}~\bibnamefont {Weih}},
		\bibinfo {author} {\bibfnamefont {M.}~\bibnamefont {Kamp}}, \bibinfo {author}
		{\bibfnamefont {J.~R.}\ \bibnamefont {Meyer}}, \bibinfo {author}
		{\bibfnamefont {C.~L.}\ \bibnamefont {Canedy}}, \bibinfo {author}
		{\bibfnamefont {C.~S.}\ \bibnamefont {Kim}}, \bibinfo {author} {\bibfnamefont
			{M.}~\bibnamefont {Kim}}, \bibinfo {author} {\bibfnamefont {W.~W.}\
			\bibnamefont {Bewley}}, \bibinfo {author} {\bibfnamefont {C.~D.}\
			\bibnamefont {Merritt}}, \bibinfo {author} {\bibfnamefont {J.}~\bibnamefont
			{Abell}},\ and\ \bibinfo {author} {\bibfnamefont {S.}~\bibnamefont
			{Höfling}},\ }\bibfield  {title} {\bibinfo {title} {Interband cascade
			lasers},\ }\href {https://doi.org/10.1088/0022-3727/48/12/123001} {\bibfield
		{journal} {\bibinfo  {journal} {Journal of Physics D: Applied Physics}\
		}\textbf {\bibinfo {volume} {48}},\ \bibinfo {pages} {123001} (\bibinfo
		{year} {2015})}\BibitemShut {NoStop}%
	\bibitem [{\citenamefont {Vurgaftman}\ \emph
		{et~al.}(2011{\natexlab{a}})\citenamefont {Vurgaftman}, \citenamefont
		{Bewley}, \citenamefont {Canedy}, \citenamefont {Kim}, \citenamefont {Kim},
		\citenamefont {Merritt}, \citenamefont {Abell}, \citenamefont {Lindle},\ and\
		\citenamefont {Meyer}}]{vurgaftman2011}%
	\BibitemOpen
	\bibfield  {author} {\bibinfo {author} {\bibfnamefont {I.}~\bibnamefont
			{Vurgaftman}}, \bibinfo {author} {\bibfnamefont {W.}~\bibnamefont {Bewley}},
		\bibinfo {author} {\bibfnamefont {C.}~\bibnamefont {Canedy}}, \bibinfo
		{author} {\bibfnamefont {C.}~\bibnamefont {Kim}}, \bibinfo {author}
		{\bibfnamefont {M.}~\bibnamefont {Kim}}, \bibinfo {author} {\bibfnamefont
			{C.}~\bibnamefont {Merritt}}, \bibinfo {author} {\bibfnamefont
			{J.}~\bibnamefont {Abell}}, \bibinfo {author} {\bibfnamefont
			{J.}~\bibnamefont {Lindle}},\ and\ \bibinfo {author} {\bibfnamefont
			{J.}~\bibnamefont {Meyer}},\ }\bibfield  {title} {\bibinfo {title}
		{Rebalancing of internally generated carriers for mid-infrared interband
			cascade lasers with very low power consumption},\ }\href
	{https://doi.org/10.1038/ncomms1595} {\bibfield  {journal} {\bibinfo
			{journal} {Nature communications}\ }\textbf {\bibinfo {volume} {2}},\
		\bibinfo {pages} {585} (\bibinfo {year} {2011}{\natexlab{a}})}\BibitemShut
	{NoStop}%
	\bibitem [{\citenamefont {Knötig}\ \emph {et~al.}(2022)\citenamefont
		{Knötig}, \citenamefont {Nauschütz}, \citenamefont {Opa\v{c}ak},
		\citenamefont {Höfling}, \citenamefont {Koeth}, \citenamefont {Weih},\ and\
		\citenamefont {Schwarz}}]{knoetig2022}%
	\BibitemOpen
	\bibfield  {author} {\bibinfo {author} {\bibfnamefont {H.}~\bibnamefont
			{Knötig}}, \bibinfo {author} {\bibfnamefont {J.}~\bibnamefont {Nauschütz}},
		\bibinfo {author} {\bibfnamefont {N.}~\bibnamefont {Opa\v{c}ak}}, \bibinfo
		{author} {\bibfnamefont {S.}~\bibnamefont {Höfling}}, \bibinfo {author}
		{\bibfnamefont {J.}~\bibnamefont {Koeth}}, \bibinfo {author} {\bibfnamefont
			{R.}~\bibnamefont {Weih}},\ and\ \bibinfo {author} {\bibfnamefont
			{B.}~\bibnamefont {Schwarz}},\ }\bibfield  {title} {\bibinfo {title}
		{Mitigating valence intersubband absorption in interband cascade lasers},\
	}\href {https://doi.org/https://doi.org/10.1002/lpor.202200156} {\bibfield
		{journal} {\bibinfo  {journal} {Laser \& Photonics Reviews}\ ,\ \bibinfo
			{pages} {2200156}} (\bibinfo {year} {2022})}\BibitemShut {NoStop}%
	\bibitem [{\citenamefont {Bader}\ \emph {et~al.}(2021)\citenamefont {Bader},
		\citenamefont {Steinbrecher}, \citenamefont {Rothmayr}, \citenamefont
		{Rawal}, \citenamefont {Hartmann}, \citenamefont {Pfenning},\ and\
		\citenamefont {Höfling}}]{bader2021}%
	\BibitemOpen
	\bibfield  {author} {\bibinfo {author} {\bibfnamefont {A.}~\bibnamefont
			{Bader}}, \bibinfo {author} {\bibfnamefont {L.}~\bibnamefont {Steinbrecher}},
		\bibinfo {author} {\bibfnamefont {F.}~\bibnamefont {Rothmayr}}, \bibinfo
		{author} {\bibfnamefont {Y.}~\bibnamefont {Rawal}}, \bibinfo {author}
		{\bibfnamefont {F.}~\bibnamefont {Hartmann}}, \bibinfo {author}
		{\bibfnamefont {A.}~\bibnamefont {Pfenning}},\ and\ \bibinfo {author}
		{\bibfnamefont {S.}~\bibnamefont {Höfling}},\ }\bibfield  {title} {\bibinfo
		{title} {{III-V semiconductor mid-infrared interband cascade light emitters
				and detectors}},\ }in\ \href {https://doi.org/10.1117/12.2599140} {\emph
		{\bibinfo {booktitle} {Infrared Remote Sensing and Instrumentation XXIX}}},\
	Vol.\ \bibinfo {volume} {11830},\ \bibinfo {editor} {edited by\ \bibinfo
		{editor} {\bibfnamefont {M.}~\bibnamefont {Strojnik}}},\ \bibinfo
	{organization} {International Society for Optics and Photonics}\ (\bibinfo
	{publisher} {SPIE},\ \bibinfo {year} {2021})\ pp.\ \bibinfo {pages} {113 --
		122}\BibitemShut {NoStop}%
	\bibitem [{\citenamefont {Bewley}\ \emph {et~al.}(2012)\citenamefont {Bewley},
		\citenamefont {Canedy}, \citenamefont {Kim}, \citenamefont {Kim},
		\citenamefont {Merritt}, \citenamefont {Abell}, \citenamefont {Vurgaftman},\
		and\ \citenamefont {Meyer}}]{bewley2012}%
	\BibitemOpen
	\bibfield  {author} {\bibinfo {author} {\bibfnamefont {W.~W.}\ \bibnamefont
			{Bewley}}, \bibinfo {author} {\bibfnamefont {C.~L.}\ \bibnamefont {Canedy}},
		\bibinfo {author} {\bibfnamefont {C.~S.}\ \bibnamefont {Kim}}, \bibinfo
		{author} {\bibfnamefont {M.}~\bibnamefont {Kim}}, \bibinfo {author}
		{\bibfnamefont {C.~D.}\ \bibnamefont {Merritt}}, \bibinfo {author}
		{\bibfnamefont {J.}~\bibnamefont {Abell}}, \bibinfo {author} {\bibfnamefont
			{I.}~\bibnamefont {Vurgaftman}},\ and\ \bibinfo {author} {\bibfnamefont
			{J.~R.}\ \bibnamefont {Meyer}},\ }\bibfield  {title} {\bibinfo {title}
		{Continuous-wave interband cascade lasers operating above room temperature at
			$\lambda$ $=$ 4.7-5.6 $\mu$m},\ }\href {https://doi.org/10.1364/OE.20.003235}
	{\bibfield  {journal} {\bibinfo  {journal} {Opt. Express}\ }\textbf {\bibinfo
			{volume} {20}},\ \bibinfo {pages} {3235} (\bibinfo {year}
		{2012})}\BibitemShut {NoStop}%
	\bibitem [{\citenamefont {Scheuermann}\ \emph {et~al.}(2017)\citenamefont
		{Scheuermann}, \citenamefont {Weih}, \citenamefont {Becker}, \citenamefont
		{Fischer}, \citenamefont {Koeth},\ and\ \citenamefont
		{Höfling}}]{scheuermann2017}%
	\BibitemOpen
	\bibfield  {author} {\bibinfo {author} {\bibfnamefont {J.}~\bibnamefont
			{Scheuermann}}, \bibinfo {author} {\bibfnamefont {R.}~\bibnamefont {Weih}},
		\bibinfo {author} {\bibfnamefont {S.}~\bibnamefont {Becker}}, \bibinfo
		{author} {\bibfnamefont {M.}~\bibnamefont {Fischer}}, \bibinfo {author}
		{\bibfnamefont {J.}~\bibnamefont {Koeth}},\ and\ \bibinfo {author}
		{\bibfnamefont {S.}~\bibnamefont {Höfling}},\ }\bibfield  {title} {\bibinfo
		{title} {{Single-mode interband cascade laser multiemitter structure for
				two-wavelength absorption spectroscopy}},\ }\href
	{https://doi.org/10.1117/1.OE.57.1.011008} {\bibfield  {journal} {\bibinfo
			{journal} {Optical Engineering}\ }\textbf {\bibinfo {volume} {57}},\ \bibinfo
		{pages} {1 } (\bibinfo {year} {2017})}\BibitemShut {NoStop}%
	\bibitem [{\citenamefont {von Edlinger}\ \emph {et~al.}(2014)\citenamefont {von
			Edlinger}, \citenamefont {Scheuermann}, \citenamefont {Weih}, \citenamefont
		{Zimmermann}, \citenamefont {Nähle}, \citenamefont {Fischer}, \citenamefont
		{Koeth}, \citenamefont {Höfling},\ and\ \citenamefont
		{Kamp}}]{edlinger2014}%
	\BibitemOpen
	\bibfield  {author} {\bibinfo {author} {\bibfnamefont {M.}~\bibnamefont {von
				Edlinger}}, \bibinfo {author} {\bibfnamefont {J.}~\bibnamefont
			{Scheuermann}}, \bibinfo {author} {\bibfnamefont {R.}~\bibnamefont {Weih}},
		\bibinfo {author} {\bibfnamefont {C.}~\bibnamefont {Zimmermann}}, \bibinfo
		{author} {\bibfnamefont {L.}~\bibnamefont {Nähle}}, \bibinfo {author}
		{\bibfnamefont {M.}~\bibnamefont {Fischer}}, \bibinfo {author} {\bibfnamefont
			{J.}~\bibnamefont {Koeth}}, \bibinfo {author} {\bibfnamefont
			{S.}~\bibnamefont {Höfling}},\ and\ \bibinfo {author} {\bibfnamefont
			{M.}~\bibnamefont {Kamp}},\ }\bibfield  {title} {\bibinfo {title} {Monomode
			interband cascade lasers at 5.2\,\textmu${\rm m}$ for nitric oxide sensing},\
	}\href {https://doi.org/10.1109/LPT.2013.2297447} {\bibfield  {journal}
		{\bibinfo  {journal} {IEEE Photonics Technology Letters}\ }\textbf {\bibinfo
			{volume} {26}},\ \bibinfo {pages} {480} (\bibinfo {year} {2014})}\BibitemShut
	{NoStop}%
	\bibitem [{\citenamefont {Yang}\ \emph {et~al.}(2019)\citenamefont {Yang},
		\citenamefont {Li}, \citenamefont {Huang}, \citenamefont {Rassel},
		\citenamefont {Gupta}, \citenamefont {Bezinger}, \citenamefont {Wu},
		\citenamefont {Razavipour},\ and\ \citenamefont {Aers}}]{yang2019}%
	\BibitemOpen
	\bibfield  {author} {\bibinfo {author} {\bibfnamefont {R.~Q.}\ \bibnamefont
			{Yang}}, \bibinfo {author} {\bibfnamefont {L.}~\bibnamefont {Li}}, \bibinfo
		{author} {\bibfnamefont {W.}~\bibnamefont {Huang}}, \bibinfo {author}
		{\bibfnamefont {S.~M.~S.}\ \bibnamefont {Rassel}}, \bibinfo {author}
		{\bibfnamefont {J.~A.}\ \bibnamefont {Gupta}}, \bibinfo {author}
		{\bibfnamefont {A.}~\bibnamefont {Bezinger}}, \bibinfo {author}
		{\bibfnamefont {X.}~\bibnamefont {Wu}}, \bibinfo {author} {\bibfnamefont
			{S.~G.}\ \bibnamefont {Razavipour}},\ and\ \bibinfo {author} {\bibfnamefont
			{G.~C.}\ \bibnamefont {Aers}},\ }\bibfield  {title} {\bibinfo {title}
		{Inas-based interband cascade lasers},\ }\href
	{https://doi.org/10.1109/JSTQE.2019.2916923} {\bibfield  {journal} {\bibinfo
			{journal} {IEEE Journal of Selected Topics in Quantum Electronics}\ }\textbf
		{\bibinfo {volume} {25}},\ \bibinfo {pages} {1} (\bibinfo {year}
		{2019})}\BibitemShut {NoStop}%
	\bibitem [{\citenamefont {Rassel}\ \emph {et~al.}(2017)\citenamefont {Rassel},
		\citenamefont {Li}, \citenamefont {Li}, \citenamefont {Yang}, \citenamefont
		{Gupta}, \citenamefont {Wu},\ and\ \citenamefont {Aers}}]{rassel2017}%
	\BibitemOpen
	\bibfield  {author} {\bibinfo {author} {\bibfnamefont {S.~M.~S.}\
			\bibnamefont {Rassel}}, \bibinfo {author} {\bibfnamefont {L.}~\bibnamefont
			{Li}}, \bibinfo {author} {\bibfnamefont {Y.}~\bibnamefont {Li}}, \bibinfo
		{author} {\bibfnamefont {R.~Q.}\ \bibnamefont {Yang}}, \bibinfo {author}
		{\bibfnamefont {J.~A.}\ \bibnamefont {Gupta}}, \bibinfo {author}
		{\bibfnamefont {X.}~\bibnamefont {Wu}},\ and\ \bibinfo {author}
		{\bibfnamefont {G.~C.}\ \bibnamefont {Aers}},\ }\bibfield  {title} {\bibinfo
		{title} {{High-temperature and low-threshold interband cascade lasers at
				wavelengths longer than 6 \textmu m}},\ }\href
	{https://doi.org/10.1117/1.OE.57.1.011021} {\bibfield  {journal} {\bibinfo
			{journal} {Optical Engineering}\ }\textbf {\bibinfo {volume} {57}},\ \bibinfo
		{pages} {1 } (\bibinfo {year} {2017})}\BibitemShut {NoStop}%
	\bibitem [{\citenamefont {Dallner}\ \emph {et~al.}(2013)\citenamefont
		{Dallner}, \citenamefont {H{\"o}fling},\ and\ \citenamefont
		{Kamp}}]{dallner2013}%
	\BibitemOpen
	\bibfield  {author} {\bibinfo {author} {\bibfnamefont {M.}~\bibnamefont
			{Dallner}}, \bibinfo {author} {\bibfnamefont {S.}~\bibnamefont
			{H{\"o}fling}},\ and\ \bibinfo {author} {\bibfnamefont {M.}~\bibnamefont
			{Kamp}},\ }\bibfield  {title} {\bibinfo {title} {Room-temperature operation
			of inas-based interband-cascade-lasers beyond 6 $\mu$m},\ }\href@noop {}
	{\bibfield  {journal} {\bibinfo  {journal} {Electronics Letters}\ }\textbf
		{\bibinfo {volume} {49}},\ \bibinfo {pages} {286} (\bibinfo {year}
		{2013})}\BibitemShut {NoStop}%
	\bibitem [{\citenamefont {Massengale}\ \emph {et~al.}(2022)\citenamefont
		{Massengale}, \citenamefont {Shen}, \citenamefont {Yang}, \citenamefont
		{Hawkins},\ and\ \citenamefont {Klem}}]{massengale2022}%
	\BibitemOpen
	\bibfield  {author} {\bibinfo {author} {\bibfnamefont {J.~A.}\ \bibnamefont
			{Massengale}}, \bibinfo {author} {\bibfnamefont {Y.}~\bibnamefont {Shen}},
		\bibinfo {author} {\bibfnamefont {R.~Q.}\ \bibnamefont {Yang}}, \bibinfo
		{author} {\bibfnamefont {S.~D.}\ \bibnamefont {Hawkins}},\ and\ \bibinfo
		{author} {\bibfnamefont {J.~F.}\ \bibnamefont {Klem}},\ }\bibfield  {title}
	{\bibinfo {title} {Long wavelength interband cascade lasers},\ }\bibfield
	{journal} {\bibinfo  {journal} {Applied Physics Letters}\ }\textbf {\bibinfo
		{volume} {120}},\ \href {https://doi.org/10.1063/5.0084565}
	{10.1063/5.0084565} (\bibinfo {year} {2022})\BibitemShut {NoStop}%
	\bibitem [{\citenamefont {{Li}}\ \emph {et~al.}(2015)\citenamefont {{Li}},
		\citenamefont {{Ye}}, \citenamefont {{Jiang}}, \citenamefont {{Yang}},
		\citenamefont {{Keay}}, \citenamefont {{Mishima}}, \citenamefont {{Santos}},\
		and\ \citenamefont {{Johnson}}}]{li2015}%
	\BibitemOpen
	\bibfield  {author} {\bibinfo {author} {\bibfnamefont {L.}~\bibnamefont
			{{Li}}}, \bibinfo {author} {\bibfnamefont {H.}~\bibnamefont {{Ye}}}, \bibinfo
		{author} {\bibfnamefont {Y.}~\bibnamefont {{Jiang}}}, \bibinfo {author}
		{\bibfnamefont {R.~Q.}\ \bibnamefont {{Yang}}}, \bibinfo {author}
		{\bibfnamefont {J.~C.}\ \bibnamefont {{Keay}}}, \bibinfo {author}
		{\bibfnamefont {T.~D.}\ \bibnamefont {{Mishima}}}, \bibinfo {author}
		{\bibfnamefont {M.~B.}\ \bibnamefont {{Santos}}},\ and\ \bibinfo {author}
		{\bibfnamefont {M.~B.}\ \bibnamefont {{Johnson}}},\ }\bibfield  {title}
	{\bibinfo {title} {{MBE-grown long-wavelength interband cascade lasers on
				InAs substrates}},\ }\href {https://doi.org/10.1016/j.jcrysgro.2015.02.016}
	{\bibfield  {journal} {\bibinfo  {journal} {Journal of Crystal Growth}\
		}\textbf {\bibinfo {volume} {425}},\ \bibinfo {pages} {369} (\bibinfo {year}
		{2015})}\BibitemShut {NoStop}%
	\bibitem [{\citenamefont {Cheng}\ \emph {et~al.}(2022)\citenamefont {Cheng},
		\citenamefont {Zhang}, \citenamefont {Sun}, \citenamefont {Zhuo},
		\citenamefont {Zhai}, \citenamefont {Liu}, \citenamefont {Wang},
		\citenamefont {Liu},\ and\ \citenamefont {Liu}}]{cheng2022}%
	\BibitemOpen
	\bibfield  {author} {\bibinfo {author} {\bibfnamefont {F.}~\bibnamefont
			{Cheng}}, \bibinfo {author} {\bibfnamefont {J.}~\bibnamefont {Zhang}},
		\bibinfo {author} {\bibfnamefont {Y.}~\bibnamefont {Sun}}, \bibinfo {author}
		{\bibfnamefont {N.}~\bibnamefont {Zhuo}}, \bibinfo {author} {\bibfnamefont
			{S.}~\bibnamefont {Zhai}}, \bibinfo {author} {\bibfnamefont {J.}~\bibnamefont
			{Liu}}, \bibinfo {author} {\bibfnamefont {L.}~\bibnamefont {Wang}}, \bibinfo
		{author} {\bibfnamefont {S.}~\bibnamefont {Liu}},\ and\ \bibinfo {author}
		{\bibfnamefont {F.}~\bibnamefont {Liu}},\ }\bibfield  {title} {\bibinfo
		{title} {High performance distributed feedback quantum cascade laser emitting
			at $\lambda$ $\sim$6.12um},\ }\href {https://doi.org/10.1364/OE.450234}
	{\bibfield  {journal} {\bibinfo  {journal} {Opt. Express}\ }\textbf {\bibinfo
			{volume} {30}},\ \bibinfo {pages} {5848} (\bibinfo {year}
		{2022})}\BibitemShut {NoStop}%
	\bibitem [{\citenamefont {Mawst}\ and\ \citenamefont
		{Botez}(2022)}]{mawst2022}%
	\BibitemOpen
	\bibfield  {author} {\bibinfo {author} {\bibfnamefont {L.~J.}\ \bibnamefont
			{Mawst}}\ and\ \bibinfo {author} {\bibfnamefont {D.}~\bibnamefont {Botez}},\
	}\bibfield  {title} {\bibinfo {title} {High-power mid-infrared ($\lambda
			\approx$ 3-6\textmu m) quantum cascade lasers},\ }\href
	{https://doi.org/10.1109/JPHOT.2021.3132261} {\bibfield  {journal} {\bibinfo
			{journal} {IEEE Photonics Journal}\ }\textbf {\bibinfo {volume} {14}},\
		\bibinfo {pages} {1} (\bibinfo {year} {2022})}\BibitemShut {NoStop}%
	\bibitem [{\citenamefont {Razeghi}\ \emph {et~al.}(2003)\citenamefont
		{Razeghi}, \citenamefont {Slivken}, \citenamefont {Yu}, \citenamefont
		{Evans},\ and\ \citenamefont {David}}]{razeghi2003}%
	\BibitemOpen
	\bibfield  {author} {\bibinfo {author} {\bibfnamefont {M.}~\bibnamefont
			{Razeghi}}, \bibinfo {author} {\bibfnamefont {S.}~\bibnamefont {Slivken}},
		\bibinfo {author} {\bibfnamefont {J.}~\bibnamefont {Yu}}, \bibinfo {author}
		{\bibfnamefont {A.}~\bibnamefont {Evans}},\ and\ \bibinfo {author}
		{\bibfnamefont {J.}~\bibnamefont {David}},\ }\bibfield  {title} {\bibinfo
		{title} {High performance quantum cascade lasers at $\lambda$$\approx$
			6$\mu$m},\ }\href
	{https://doi.org/https://doi.org/10.1016/S0026-2692(03)00030-2} {\bibfield
		{journal} {\bibinfo  {journal} {Microelectronics Journal}\ }\textbf {\bibinfo
			{volume} {34}},\ \bibinfo {pages} {383} (\bibinfo {year} {2003})},\ \bibinfo
	{note} {the Fourth International Conference on Low Dimensional Structures and
		Devices}\BibitemShut {NoStop}%
	\bibitem [{\citenamefont {Canedy}\ \emph {et~al.}(2017)\citenamefont {Canedy},
		\citenamefont {Warren}, \citenamefont {Merritt}, \citenamefont {Bewley},
		\citenamefont {Kim}, \citenamefont {Kim}, \citenamefont {Vurgaftman},\ and\
		\citenamefont {Meyer}}]{canedy2017}%
	\BibitemOpen
	\bibfield  {author} {\bibinfo {author} {\bibfnamefont {C.~L.}\ \bibnamefont
			{Canedy}}, \bibinfo {author} {\bibfnamefont {M.~V.}\ \bibnamefont {Warren}},
		\bibinfo {author} {\bibfnamefont {C.~D.}\ \bibnamefont {Merritt}}, \bibinfo
		{author} {\bibfnamefont {W.~W.}\ \bibnamefont {Bewley}}, \bibinfo {author}
		{\bibfnamefont {C.~S.}\ \bibnamefont {Kim}}, \bibinfo {author} {\bibfnamefont
			{M.}~\bibnamefont {Kim}}, \bibinfo {author} {\bibfnamefont {I.}~\bibnamefont
			{Vurgaftman}},\ and\ \bibinfo {author} {\bibfnamefont {J.~R.}\ \bibnamefont
			{Meyer}},\ }\bibfield  {title} {\bibinfo {title} {{Interband cascade lasers
				with longer wavelengths}},\ }in\ \href {https://doi.org/10.1117/12.2246450}
	{\emph {\bibinfo {booktitle} {Quantum Sensing and Nano Electronics and
				Photonics XIV}}},\ Vol.\ \bibinfo {volume} {10111},\ \bibinfo {editor}
	{edited by\ \bibinfo {editor} {\bibfnamefont {M.}~\bibnamefont {Razeghi}}},\
	\bibinfo {organization} {International Society for Optics and Photonics}\
	(\bibinfo  {publisher} {SPIE},\ \bibinfo {year} {2017})\ pp.\ \bibinfo
	{pages} {80 -- 86}\BibitemShut {NoStop}%
	\bibitem [{\citenamefont {Vurgaftman}\ \emph
		{et~al.}(2011{\natexlab{b}})\citenamefont {Vurgaftman}, \citenamefont
		{Bewley}, \citenamefont {Canedy}, \citenamefont {Kim}, \citenamefont {Kim},
		\citenamefont {Lindle}, \citenamefont {Merritt}, \citenamefont {Abell},\ and\
		\citenamefont {Meyer}}]{vurgaftman2011ieee}%
	\BibitemOpen
	\bibfield  {author} {\bibinfo {author} {\bibfnamefont {I.}~\bibnamefont
			{Vurgaftman}}, \bibinfo {author} {\bibfnamefont {W.~W.}\ \bibnamefont
			{Bewley}}, \bibinfo {author} {\bibfnamefont {C.~L.}\ \bibnamefont {Canedy}},
		\bibinfo {author} {\bibfnamefont {C.~S.}\ \bibnamefont {Kim}}, \bibinfo
		{author} {\bibfnamefont {M.}~\bibnamefont {Kim}}, \bibinfo {author}
		{\bibfnamefont {J.~R.}\ \bibnamefont {Lindle}}, \bibinfo {author}
		{\bibfnamefont {C.~D.}\ \bibnamefont {Merritt}}, \bibinfo {author}
		{\bibfnamefont {J.}~\bibnamefont {Abell}},\ and\ \bibinfo {author}
		{\bibfnamefont {J.~R.}\ \bibnamefont {Meyer}},\ }\bibfield  {title} {\bibinfo
		{title} {Mid-ir type-ii interband cascade lasers},\ }\href
	{https://doi.org/10.1109/JSTQE.2011.2114331} {\bibfield  {journal} {\bibinfo
			{journal} {IEEE Journal of Selected Topics in Quantum Electronics}\ }\textbf
		{\bibinfo {volume} {17}},\ \bibinfo {pages} {1435} (\bibinfo {year}
		{2011}{\natexlab{b}})}\BibitemShut {NoStop}%
	\bibitem [{\citenamefont {Bewley}\ \emph {et~al.}(2004)\citenamefont {Bewley},
		\citenamefont {Canedy}, \citenamefont {Kim}, \citenamefont {Vurgaftman},
		\citenamefont {Kim},\ and\ \citenamefont {Meyer}}]{bewley2004}%
	\BibitemOpen
	\bibfield  {author} {\bibinfo {author} {\bibfnamefont {W.}~\bibnamefont
			{Bewley}}, \bibinfo {author} {\bibfnamefont {C.}~\bibnamefont {Canedy}},
		\bibinfo {author} {\bibfnamefont {C.}~\bibnamefont {Kim}}, \bibinfo {author}
		{\bibfnamefont {I.}~\bibnamefont {Vurgaftman}}, \bibinfo {author}
		{\bibfnamefont {M.}~\bibnamefont {Kim}},\ and\ \bibinfo {author}
		{\bibfnamefont {J.}~\bibnamefont {Meyer}},\ }\bibfield  {title} {\bibinfo
		{title} {Antimonide type-ii “w” lasers: growth studies and guided-mode
			leakage into substrate},\ }\href
	{https://doi.org/https://doi.org/10.1016/j.physe.2003.08.060} {\bibfield
		{journal} {\bibinfo  {journal} {Physica E: Low-dimensional Systems and
				Nanostructures}\ }\textbf {\bibinfo {volume} {20}},\ \bibinfo {pages} {466}
		(\bibinfo {year} {2004})},\ \bibinfo {note} {proceedings of the 11th
		International Conference on Narrow Gap Semiconductors}\BibitemShut {NoStop}%
\end{thebibliography}
\end{document}